\documentstyle[preprint,epsbox,seceq]{jpsj}

\def\GA{\raise2.5pt\hbox{$>$}\kern-8pt\lower2.5pt\hbox{$\sim$}}
\def\LA{\raise2.5pt\hbox{$<$}\kern-8pt\lower2.5pt\hbox{$\sim$}}

\def\runtitle{
Electrical Conductivity of Fermi Liquids. I. 
}
\def\runauthor{Takuya OKABE}

\title{
Electrical Conductivity of Fermi Liquids. I. \\
Many-body Effect on the Drude Weight
}

\author
{ 
Takuya {\sc Okabe}
\footnote{Present address: 
Faculty of Engineering, Gunma University, Kiryu, Gunma 376.}
\footnote{E-mail: okabe@phys.eg.gunma-u.ac.jp}
}
\inst
{
Department of Physics, Kyoto University, Kyoto 606-01
}

\recdate
{
February 12, 1998}

\abst
{
On the basis of the Fermi liquid theory,
we investigate the many-body effect on the Drude weight.
In a lattice system, 
the Drude weight $D$ 
is modified by electron-electron interaction
due to Umklapp processes,
while it is not renormalized in a Galilean invariant system.
This is explained by showing that 
the effective mass $m'$ for $D\propto n/m'$
is defined through the current, not velocity, of quasiparticle.
It is shown that the inequality $D>0$ is required for the 
stability against the uniform shift of the Fermi surface.
The result of perturbation theory 
applied for the Hubbard model 
indicates that $D$ as a function of the density $n$
is qualitatively modified around half filling $n\sim 1$
by Umklapp processes. 
}

\kword
{
Fermi liquid theory, Drude weight, many-body effect, 
Umklapp process , metal-insulator transition,
}

\begin{document}
\sloppy
%\twocolumn
\maketitle

\section{Introduction}

In comparing theory and experiment of 
highly correlated electron systems,
we must distinguish genuine many-body effects from one-body effects.
This is trivial at least 
in a theoretical investigation based on a simple model.
For example, in the Fermi liquid theory,
the thermal mass measured in a specific heat experiment
reflects the mass of quasiparticles,
which can be heavily enhanced by interaction effects.
On the other side, the effective mass $m'$ defined 
in the Drude weight, 
which characterizes 
the metallic property of the system, 
does not generally coincide with the quasiparticle mass $m^*$.
In fact it is shown that
$m'$ is not renormalized, or
it is independent of electron-electron interactions
in Galilean invariant systems.
In a lattice system, Umklapp processes make
the Drude weight depend on the electron-electron interaction and
thus the Drude weight can be used as a direct probe of
the metal-insulator transition.~\cite{rf:Kohn,rf:SS}
This problem has been studied 
particularly 
in one dimensional systems,~\cite{rf:SS,rf:ELP,rf:UKO,rf:1d}\
where it is shown that the Drude weight $D$ vanishes 
as the carrier density $n$ approaches 
half filling, i.e.,  
in the presence of the Coulomb interaction $U> 0$,
we have $D\propto |1-n|=\delta$
for the low density limit $\delta\rightarrow 0$ of doped hole.
Though the metal-insulator transition in this form
might not be realized in general,
we are interested in how this qualitative behavior 
of `doped insulator' is reconciled with 
normal Fermi liquids in two or three dimensional systems,
and we investigate and show how 
it is put in the context 
of the general theory of Fermi liquids
when Umklapp processes are effective.
In this respect, 
the effect of Umklapp processes in Fermi liquids
was discussed for the cyclotron resonance frequency $\omega_c$
by Kanki and Yamada,~\cite{rf:KY}\
where it was shown that $\omega_c$ is not renormalized 
in an isotropic system.
In a similar manner, we discuss 
that $D$ is not renormalized in an isotropic system 
and then see in what manner $D$ is modified
in a lattice system.

%In this paper, we investigate the many-body effect on
%the Drude weight $D$ on the basis of the Fermi liquid theory.

In \S\ref{FLTI} we derive the expression for $D$
on the basis of the Landau Fermi liquid theory
and show that it is calculated as
the second derivative of the energy $E_p$
of the state $|p\rangle$,
which is defined from the ground state 
by shifting the fermion distribution function $n^0_k$. 
The results obtained in \S\ref{FLTI} are derived microscopically 
in \S\ref{FLTII}, where the Fermi liquid result of $D$ is 
obtained using
\begin{equation}
D\equiv \lim_{\omega\rightarrow 0}
\pi\omega \mbox{Im}\sigma(\omega).
\label{defD}
\end{equation}
To supplement the general discussion of 
\S\S \ref{FLTI} and \ref{FLTII},
$D$ is estimated explicitly in \S\ref{PT}, 
where 
the effective mass for the Drude weight
is calculated  up to the second order in $U$
of the Hubbard model in a square lattice.
We observe that $D$ as a function of $n$
is modified qualitatively around half filling $n\sim 1$.
The last section 
%(\S\ref{sec:D}) 
contains discussions.
%, \S\ref{sec:D}.
In this article, we investigate a system
described by a single-band model of fermions at absolute zero.
We are mainly interested in 
the collisionless region where the effect of 
a quasiparticle lifetime can be neglected.
In Appendix we outline 
the derivation of one of the main results
on the basis of the finite temperature formalism.
A general theory at finite temperature will 
be presented in a subsequent paper.~\cite{rf:2}\

\section{Fermi Liquid Theory I}
\label{FLTI}

First we derive the uniform conductivity
on the basis of the Landau theory of Fermi liquids.~\cite{rf:PN}\ 
In the presence of an applied electric field $\mib{E}$,
the deviation $\delta n_{k}$
from the ground-state distribution function $n^0_{k}$,
\[
\delta n_k=n_{k}-n^0_{k},
\]
satisfies the Boltzmann equation,
\begin{eqnarray}
&&(\mib{q}\mib \cdot \mib{v}_{k}-\omega)\delta n_{k}
+\mib{q\cdot v}_{k}\delta(\mu-\varepsilon_k)
\sum_{k'}f(k,k')\delta n_{k'}\nonumber\\
&&\qquad+\mbox{i} e \mib{E\cdot v}_{k}\delta(\mu-\varepsilon_k)=0,
\end{eqnarray}
where $\varepsilon_{k}$ is energy of the quasiparticle $k$
and 
\[{ v}_{k}\equiv
 \frac{\partial \varepsilon_{k}}{\partial k}.\]
For $\mib q= 0$ we obtain 
\begin{equation}
\delta n_{k}=\frac{\mbox{i}e\mib{E\cdot v}_{k}
\delta(\mu-\varepsilon_{k})}{\omega}.
\end{equation}
The total current induced by the field is 
\begin{equation}
{\mib J}=\sum_{k}\delta n_{k}\mib j_{k}.
\end{equation}
The conductivity tensor is defined by
\begin{equation}
\sigma_{\mu\nu}(\omega)=\frac{eJ_\mu}{\Omega E_\nu},
\label{Defsig}
\end{equation}
where $\Omega$ is the total volume of the system.
Hence we find
\begin{equation}
\sigma_{\mu\nu}(\omega)=\frac{{\mbox i}e^2}{\omega}
\left(\frac{n}{m'}\right)_{\mu\nu},
\label{conductivity}
\end{equation}
where
\begin{equation}
\left(\frac{n}{m'}\right)_{\mu\nu}
=\frac{1}{\Omega}\sum_{ k}v_{k\mu}j_{k\nu}
\delta(\mu-\varepsilon_{k}).
\label{nm'FL}
\end{equation}
The Drude weight $D$, eq.~(\ref{defD}),
is given as the coefficient of the delta function $\delta(\omega)$ 
in $\mbox{Re} \sigma_{\mu\nu}(\omega)$;
%is expressed in terms of the effective mass $m'$ defined here,
%in eq.~(\ref{nm'FL})
\begin{equation}
\mbox{Re} \sigma_{\mu\nu}(\omega)=
\pi e^2\left(\frac{n}{m'}\right)_{\mu\nu}
\delta(\omega).
\label{Drude}
\end{equation}
Therefore, we obtain
\[
D=\pi e^2\left(\frac{n}{m'}\right)_{\mu\nu}.
\]
The effective mass $m'$, defined by eq.~(\ref{nm'FL}),
plays an important role in our theory.

To see that electron-electron interaction does not modify
the Drude weight in Galilean invariant systems,
eq.~(\ref{nm'FL}) is transformed as follows,
\begin{eqnarray}
\left(\frac{n}{m'}\right)_{\mu\nu}
&=&\frac{1}{\Omega}
\sum_{k}
\left(-\frac{\partial n^0_{k}}{\partial k_\mu}\right)
j_{k\nu}\label{nm'FL2p}\\
&=&\frac{1}{\Omega}\sum_{k} n^0_{k}\frac{\partial j_{k\nu}}{
\partial k_\mu}.
\label{nm'FL2}
\end{eqnarray}
Here we used 
\begin{equation}
\delta(\mu-\varepsilon_{k})=
-\frac{\partial n^0_{k}}{\partial \varepsilon_{k}},
\end{equation}
i.e., the quasiparticle distribution function $n^0_{k}$
is represented by the step function $\theta(\mu-\varepsilon_k)$
at zero temperature.
As a direct consequence of Galilean invariance, 
the relation \(j_{k\nu}=k_\nu/m\),
with the bare electron mass $m$,
is not modified by electron-electron interactions.
Therefore in this case we obtain
\begin{equation}
\left(\frac{n}{m'}\right)_{\mu\nu}
=\frac{n}{m}\delta_{\mu\nu},
\end{equation}
where 
\begin{equation}
n\equiv\frac{1}{\Omega}\sum_{k} n^0_{k},
\end{equation}
and the Drude weight
$D=\pi e^2 n/m$ is not affected by the interaction.
If we assume $\partial j_{k\mu}/\partial k_\nu=$const.
for simplicity,
from eq.~(\ref{nm'FL2}) we find
\begin{equation}
\left(\frac{1}{m'}\right)_{\mu\nu}=
\frac{\partial j_{k\mu}}{\partial k_\nu}
\neq
\left(\frac{1}{m^*}\right)_{\mu\nu}\equiv
\frac{\partial v_{k\mu}}{\partial k_\nu},
\label{1/m'ne1/m*}
\end{equation}
where $m^*$ 
represents the effective mass of the quasiparticle $k$.
Hence it is concluded that 
the mass $m'$ in the Drude weight is given 
through the current of quasiparticle,
and that $m'$, $m^*$ and the bare (crystalline) mass $m$
generally take different values.
We remark that the effective mass $m^*$ thus defined 
by $1/m^* \propto \langle v\rangle$ 
does not generally coincide with 
the thermal mass defined in the total density of states 
$m^*_{\rm th}\propto \langle v^{-1}\rangle$,
where the average is taken over the Fermi surface.
The difference between $m^*$ and $m^*_{\rm th}$ 
would be significant only in exceptional cases,
as in the vicinity of the van Hove singularity
where the former is less singular than the latter.

Now let us show that  
the right-hand side of eq.~(\ref{nm'FL})
is derived in terms of the state $|p\rangle $,
which represents the state obtained by 
displacing the ground state configuration by $p$
in the momentum space,
i.e., $|p\rangle $ is derived
from the ground state $|0\rangle $
by formally replacing the distribution function $n^0_{k}$ 
with $n^0_{k-p}$.
By construction,
the resulting state $|p\rangle $ 
carries 
a finite current ${\mib J}_{\mib p}$.~\cite{rf:comment}\
Denoting the energy of the state $|p\rangle $ as $E_p$,
we would like to derive $J_{p\mu}=\partial E_p/\partial p_\mu$
and express $(n/m')_{\mu\nu}$ as the second derivative of $E_p$
with respect to $p$.

According to the Landau Fermi liquid theory,
the current carried by the quasiparticle $k$ 
is given by~\cite{rf:PN} 
\begin{equation}
{\mib j_{k}}=
{\mib v_{k}}
+\sum_{ k'}f(k,k'){\mib v_{k'}}
\delta(\mu-\varepsilon_{k'}).
\label{jp}
\end{equation}
The spin dependence
of the interaction function $f(k,k')$ was
omitted for simplicity.
Using
\(
\delta n_{k}=n^0_{k- p}-n^0_{k}=
-{\mib p}{\mib \cdot}{\mib \nabla_{\mib k}} n^0 =
\delta(\mu-\varepsilon_{k})
{\mib p}{\mib \cdot}{\mib v_{k}}, \)
%which is valid to the first order in ${\mib p}$,
we obtain
\begin{eqnarray}
{\mib J}_{\mib p}&=&
\sum_{k}\delta n_{k}{\mib j_{k}}
=\sum_{k}
\delta(\mu-\varepsilon_{k})
{\mib p}{\mib \cdot}{\mib v_k}{\mib j_{k}}
\label{Jp2}\\
&=&\sum_{k}
\delta(\mu-\varepsilon_{k})
{\mib p}{\mib \cdot}{\mib v_{k}}
\nonumber\\&&
\times \left(
{\mib v_{k}}
+\sum_{k'}f(k,k')
\delta(\mu-\varepsilon_{k'})
{\mib v_{k'}}\right).
\label{delJ}
\end{eqnarray}
\halftext
Comparing eqs.~(\ref{nm'FL}) and (\ref{Jp2}),
we obtain
\begin{equation}
\frac{\partial J_{p\mu}}{\partial p_\nu}
=\Omega\left(\frac{n}{m'}\right)_{\mu\nu}.
\label{delJpdelp=nm'}
\end{equation}
On the other side,
to estimate the change of the total energy 
\begin{equation}
\delta E_{p}=\sum_{k}\varepsilon_{k}\delta n_{k}
+\frac{1}{2}\sum_{k,k'}f(k,k')
\delta n_{k}\delta n_{k'},
\label{functional}
\end{equation}
the deviation $\delta n_{k}$ has to be
expanded up to the second order in $ p$:
\(\delta n_{k}=
-\sum_i p_i \partial_{k_i}n^0
+\sum_{i,j}p_i p_j \partial_{k_i}\partial_{k_j}n^0/2\).
Then we have
\begin{eqnarray}
\frac{\partial E_p}{\partial p_\mu}
&=&\sum_{k}\varepsilon_{k}
{\mib p}{\mib \cdot}{\mib \nabla_{\mib k}}\partial_{k_\mu}n^0
\nonumber\\
&+&\sum_{k,k'}f(k,k')
\delta(\mu-\varepsilon_{k})
\delta(\mu-\varepsilon_{k'})
{\mib p}{\mib \cdot}{\mib v}_{k}v_{k'\mu},
\nonumber\\
\label{delEp}
\end{eqnarray}
where the symmetry relation 
\( f(k,k')=f(k',k)\) was used
in the second term.
The first term is put into 
\begin{equation}
-\sum_{k}\left(\partial_{k_\mu}
\varepsilon_{k}\right)
{\mib p}{\mib \cdot}{\mib \nabla_{\mib k}}n^0
=\sum_{k}
\delta(\mu-\varepsilon_{k})
{\mib p}{\mib \cdot}\mib v_{k} v_{k\mu},
\end{equation}
where we used \(\sum_{k}\partial_{k_\mu}(
\varepsilon_{k}
{\mib p}{\mib \cdot}{\mib \nabla_{\mib k}} n^0)=0\).
Hence we have
\begin{eqnarray}
\frac{\partial E_p}{\partial p_\mu}&=&
\sum_{k}
\delta(\mu-\varepsilon_{k})
{\mib p}{\mib \cdot}\mib v_{k} v_{k\mu}
\nonumber\\
&+&\sum_{k,k'}f(k,k')
\delta(\mu-\varepsilon_{k})
\delta(\mu-\varepsilon_{k'})
{\mib p}{\mib \cdot}{\mib v_{k}}v_{k'\mu}.
\nonumber\\
\label{delE}
\end{eqnarray}
As a result, from eqs.~(\ref{delJ}) and (\ref{delE})
we obtain
\begin{equation}
J_{p\mu}=\frac{\partial E_p}{\partial p_\mu},
\end{equation}
and using eq.~(\ref{delJpdelp=nm'})
we conclude
\begin{equation}
\frac{\partial J_{p\mu}}{\partial p_\nu}
=\frac{\partial^2 E_p}{\partial p_\mu\partial p_\nu}
=\Omega\left(\frac{n}{m'}\right)_{\mu\nu}.
\label{J=delEdelp}
\end{equation}
As the assumption of an isotropic system,
to regard $f(k,k')$ just as a function of 
$\cos\theta=\mib{k\cdot k'}/kk'$, is not used
in the above derivation,
the result (\ref{J=delEdelp})
is generally applied even for an anisotropic system.~\cite{rf:comment1}\
%In fact, the relation (\ref{J=delEdelp})
%is generally obtained by treating $\delta n_{\mib p'}$,
%which is equivalent to the chemical potential shift,
%as a perturbation applied to the ground state.

%\section{Microscopic Theory}
%\label{MT}
\section{Fermi Liquid Theory II}
\label{FLTII}
In order to describe the state $|p\rangle$ microscopically,
we introduce the Green's function
\begin{equation}
G'^0_p(k)=\frac{1}{\omega-\varepsilon_{k}+\mu
+\mbox{i sign}(\varepsilon_{k-p}-\mu)}.
\label{G'0p}
\end{equation}
We use the abbreviated notation 
such as $k=(\mib{k},\omega)$ and
$k'=(\mib{k'},\omega')$.
In eq.~(\ref{G'0p}),  $\varepsilon_k$ represents 
energy of the bare particle $k$.
In the following,
the energy of the quasiparticle $k$
will be denoted by $\varepsilon^*_k$.
To calculate $E_p$ of the state $|p\rangle$, 
the Green's function (\ref{G'0p}) is to be used
%a modified form of the bare Green's function,
instead of
\[
G^0(k)=\frac{1}{\omega-\varepsilon_{k}+\mu
+\mbox{i sign}(\varepsilon_{k}-\mu)},
\]
which is usually used to describe 
the ground state.~\cite{rf:comment3}\
In place of eq.~(\ref{G'0p}), 
we may equivalently use
\begin{equation}
G^0_p(k)=\frac{1}{\omega-\varepsilon_{k+p}+\mu
+\mbox{i sign}(\varepsilon_{k}-\mu)
}.\label{G0p}
\end{equation}
This Green's function
%(\ref{G0p})
describes the system
which has the same Fermi surface as for the ground state,
but with the shifted one-body energy $\varepsilon_{k+p}$
instead of $\varepsilon_{k}$.
%Note that there is a relation $G'^0_p(k)=G^0_p(k-p)$.
%since there is a relation \(G'^0_p(k)=G^0_p(k+p)\).
For example, the one-body part of the total energy
of $| p\rangle$
is expressed in terms of $G'^0_p(k)=G^0_p(k-p)$,
\begin{eqnarray}
E_p^0&=&\sum_{k} \varepsilon_{k}
\langle \hat{c}^\dagger_{k}\hat{c}_{k}\rangle_p^0
=
-\mbox i%\Omega
\int\frac{{\mbox d}^4k}{(2\pi)^4}
\varepsilon_{k} G'^0_p(k)\nonumber\\
&=&
-\mbox i%\Omega
\int\frac{{\mbox d}^4k}{(2\pi)^4}
\varepsilon_{k+p} G^0_p(k)\nonumber\\
&=&
-\mbox i%\Omega
\int\frac{{\mbox d}^4k}{(2\pi)^4}
\varepsilon_{k+p} G^0(k)
\label{vk+pG0}\\
&=&\sum_{k} \varepsilon_{k+p}
\langle \hat{c}^\dagger_{k}\hat{c}_{k}\rangle^0.\nonumber
\end{eqnarray}

%%%%%%%%%%%%%%%%%%%%%%%%
To evaluate $\partial E_p/\partial p$,
the $p$-dependence in the argument of $G^0_p(k-p)$
can be neglected by virtue of 
quasimomentum conservation at each interaction point, or
 we can replace $G'^0_p(k)$ in $E_p$
by $G^0_p(k)$ without loss of generality. 
Physically, this is related to the fact that
replacing the distribution function $n_{k}$ by $n_{k-p}$
in the expression for $E_p$ is equivalent to 
replacing $\varepsilon_{k}$ by $\varepsilon_{k+p}$.
%with fixed $n_{k}$.
This point will be seen below in \S\ref{PT} for a simple case.
As a result,
for the energy correction $\Delta E_p \equiv E_p-E_p^0$
caused by electron-electron interaction,
we obtain the equation
\begin{equation}
\frac{\partial\Delta E_p}{\partial p_\mu}
=-\mbox{i}%\Omega
\int\frac{{\mbox d}^4k}{(2\pi)^4}
\frac{\partial G^0_p(k)}{\partial p_\mu}\Sigma'_p(k),
\end{equation}
where %$\Omega$ represents the volume of the system, and
$\Sigma'_p(k)$ is the total self energy
including improper as well as proper parts.
In terms of the identity
\begin{equation}
\frac{\partial G^0_p(k)}{\partial p_\mu}= v_{k+p\mu}G^0_p(k)^2,
\end{equation}
we have
\begin{eqnarray}
\lefteqn{\frac{\partial \Delta E_p}{\partial p_\mu}
=-\mbox{i}%\Omega
\int\frac{{\mbox d}^4k}{(2\pi)^4}
v_{k+p\mu}G^0_p(k)^2\Sigma'_p(k)
\label{pEpp0}}\\
&=&-\mbox{i}%\Omega
\int\frac{{\mbox d}^4k}{(2\pi)^4}
v_{k+p\mu}G^0_p(k)\Sigma_p(k)G_p(k),
\label{pEpp1}
\end{eqnarray}
where we used the relation
$\Sigma'_p(k)G^0_p(k)=\Sigma_p(k)G_p(k)$
for the proper self-energy $\Sigma_p(k)$ and
the dressed Green's function defined  by
\begin{equation}
G_p(k)^{-1}=G^0_p(k)^{-1}-\Sigma_p(k).
\label{Gpp'}
\end{equation}
Thus, together with
\begin{equation}
\frac{\partial E^0_p}{\partial p_\mu}
=-\mbox i%\Omega
\int\frac{{\mbox d}^4k}{(2\pi)^4}
v_{k+p\mu} G^0_p(k),
%\quad (=J^0_p),
\end{equation}
eq.~(\ref{pEpp1}) leads to
\begin{eqnarray}
\frac{\partial E_p}{\partial p_\mu}
&=&-\mbox{i}%\Omega
\int\frac{{\mbox d}^4k}{(2\pi)^4}
v_{k+p\mu}G_p(k),
\label{pEpp}
\end{eqnarray}
where we used
\begin{equation}
G_p(k)=G^0_p(k)+G^0_p(k)\Sigma_p(k)G_p(k).
\end{equation}
The total current of the state $|p\rangle$
is just given by eq.~(\ref{pEpp}),
since we find
\begin{eqnarray}
J_p&\equiv&\sum_{k} v_{k}
\langle \hat{c}^\dagger_{k}\hat{c}_{k}\rangle_p
=-\mbox i%\Omega
\int\frac{{\mbox d}^4k}{(2\pi)^4}
v_{k} G'_p(k)\nonumber\\
&=&-\mbox i%\Omega
\int\frac{{\mbox d}^4k}{(2\pi)^4}
v_{k+p} G_p(k).
\label{Jp}
\end{eqnarray}
Therefore we obtain
\begin{equation}
J_{p\mu}=\frac{\partial E_p}{\partial p_\mu}.
\label{Jp=delEp}
\end{equation}

Now we derive the other results of the previous section
on the basis of the microscopic Fermi liquid theory
using well-known Fermi liquid relations.~\cite{rf:Noz,rf:AGD}\
To calculate a linear response
to the vector potential $\mib A$,
the total current operator is defined by
\begin{equation}
\hat J_\mu=
\hat v_\mu-e
\sum_\nu\hat{\varepsilon}''_{\mu\nu}
 A_\nu,
\end{equation}
where
\begin{equation}
\hat{v}_\mu\equiv
\sum_k v_{k\mu} \hat{c}^\dagger_k\hat{c}_k,\qquad
v_{k\mu}=\frac{\partial \varepsilon_k}{\partial k_\mu},
\label{hatv}
\end{equation}
and
\begin{equation}
\hat{\varepsilon}''_{\mu\nu}\equiv
\sum_k \frac{\partial^2\varepsilon_k}{\partial k_\mu\partial k_\nu}
 \hat{c}^\dagger_k\hat{c}_k
\label{hatep''}.
\end{equation}
Then the dynamical conductivity is given by
\fulltext
\begin{equation}
\sigma_{\mu\nu}(\omega)=\frac{e^2}{\omega+\mbox i0}
\left(
%-\frac{1}{2\pi\mbox i\Omega}
\mbox{i}K_{\mu\nu}(\omega+\mbox i 0)+\frac{1}{\Omega}
\int\frac{\mbox d^4 k}{(2\pi)^4}
%\frac{1}{\Omega}\sum_{k}
\frac{\partial^2 \varepsilon_{k}}{\partial k_\mu \partial k_\nu}  
G(k)e^{+\rm{ i} \omega 0}
\right),
\label{sigmaKF}
\end{equation}
%Hereafter we use the abbreviated notation such as
%$p=(\mib k, \omega)$ and $k=(\mib k, \omega')$.
where
$K_{\mu\nu}(\omega)$, defined by
\begin{equation}
K_{\mu\nu}(\omega)=-\frac{\mbox i}{\Omega}
\int^\infty_{-\infty}
\langle 0|\mbox T \hat v_\mu(t)\hat v_\nu |0\rangle
\exp(\mbox i\omega t)\mbox d t,
\end{equation}
corresponds to the uniform limit $k\rightarrow 0$ of 
the current correlation function $K_{\mu\nu}(k)$.
For the Green's function $G(k)$ of the ground state,
we omit the subscript $p$ $(=0)$ for simplicity.
In terms of the vertex function $\Lambda_\nu(k';k)$,
$K_{\mu\nu}(k)$ is given by
\begin{equation}
K_{\mu\nu}(k)=-\frac{\mbox{i}}{\Omega}
\int\frac{\mbox d^4 k'}{(2\pi)^4}
%\frac{1}{\Omega}\sum_{k}
v_{k\mu} G(k'+k/2)G(k'-k/2)\Lambda_\nu(k';k).
\label{Kmunu}
\end{equation}
The vertex function satisfies the following equation;
\begin{equation}
\Lambda_\mu(k';k)= v_{k'\mu}
-\mbox i
\int\frac{\mbox d^4 k''}{(2\pi)^4}
\Gamma^{(0)}( k', k'')
 G(k''+k/2)G(k''-k/2)\Lambda_\mu(k'';k),
\label{Lambda}
\end{equation}
where
$\Gamma^{(0)}( k, k')$ is the irreducible
four-point vertex function.
As is well known, in the limit $k=(\mib{k},\omega)\rightarrow 0$,
which corresponds to the $k$-$\omega$ region in the vicinity
of the Fermi surface,
the product of the Green's functions
$G(k'+k/2)G(k'-k/2)$ does not behave regularly
but it is written
%To treat a singularity present in this limit, let us define
\fulltext
\begin{equation}
G(k'+k/2)G(k'-k/2)
=G(k')^2+2\pi \mbox i z_{k'}^2
\frac{\mib{k\cdot v^*_{k' }}}
{\omega-\mib{k\cdot v^*_{k' }}}
\delta(\mu-\varepsilon^*_{k'})
\delta(\omega').
%%
%%\left\{G(k')^2\right\}^r=
%\lim_{{\scriptstyle p\rightarrow 0}\atop{\scriptstyle r=k/\omega}}
%G(k'+p/2)G(k'-p/2)
%=G(k')^2+2\pi \mbox i z_{k'}^2
%\frac{r\cdot v^*_{k' }}
%{1-r\cdot v^*_{k' }}
%\delta(\mu-\varepsilon^*_{k'})
%\delta(\mu-\omega').
\label{G2r}
\end{equation}
The first term $G(k')^2$ represents a regular part of
the left-hand side.
In particular, we are interested in
the $\omega$-limit ($k/\omega=0$) and the $k$-limit 
($k/\omega=\infty$)
of eq.~(\ref{G2r}),
for which we obtain the relation
\begin{equation}
\left\{G(k')^2\right\}^\omega-\left\{G(k')^2\right\}^k
=2\pi \mbox i z_{k'}^2
\delta(\mu-\varepsilon^*_{k'})
\delta(\omega').
\label{difG2}
\end{equation}
\halftext
%In the $\omega$-limit ($r=k/\omega=0$),
%the limit $\omega\rightarrow 0$ is taken for $k=0$,
%while the $k$-limit ($r=\infty$) represents
%the limit $k\rightarrow 0$ with $\omega=0$.
In eq.~(\ref{G2r}),
$v^*_k$ represents the velocity of the quasiparticle $k$,
\begin{equation}
v^*_k=\frac{\partial \varepsilon^*_k}{\partial k}.
\label{QPv}\end{equation}
The quasiparticle energy $\varepsilon^*_k$
is given as a pole of the Green's function,
\begin{equation}
G^{-1}(k,\varepsilon^*_k-\mu)=\varepsilon^*_k-
\varepsilon_k-\Sigma(k,\varepsilon^*_k-\mu)=0,
\label{G-1=0}
\end{equation}
where the many-body effect in the one-particle spectrum
is embodied in the selfenergy $\Sigma(k,\omega)$.
The renormalization factor $z_{k}$
is defined by
\begin{equation}
z_k^{-1}=\left.1-\frac{\partial \Sigma(k,\omega)}{\partial \omega}
\right|_{\omega=0}.
\end{equation}
Differentiating eq.~(\ref{G-1=0}) with respect to $k$,
we have
\begin{equation}
v^*_k\left(
\left.
  1-\frac{\partial \Sigma(k,\omega)}{\partial \omega}
\right|_{\omega=0}
\right)
-v_k
-\left.
\frac{\partial \Sigma(k,\omega)}{\partial k}
\right|_{\omega=0}
=0,
\end{equation}
thus we find
\begin{equation}
v^*_k=z_k\left(
v_k+\frac{\partial \Sigma(k,0)}{\partial k}
\right).
\label{v*}
\end{equation}
%Because of $\mbox{Im}\Sigma(k,0)=0$,
%the quasiparticle energy $\varepsilon^*_k$ is a real quantity
%at the Fermi level and $\Sigma(k,\omega)$ in the above expressions
%represents the real part of the selfenergy.

Next we consider the derivative of the selfenergy,
\begin{equation}
\frac{\partial \Sigma(k,\omega)}{\partial k_\mu}
=-\mbox i
\int\frac{\mbox d^4 k'}{(2\pi)^4}
\Gamma^{(0)}(k,k')\frac{\partial}{\partial k'_\mu}G(k',\omega'),
\label{pSpk'}
\end{equation}
for which, using
\begin{equation}
\frac{\partial}{\partial k'_\mu}G(k',\omega')=
\left\{G(k')^2\right\}^k%\infty
\left(
v_{k'\mu}+\frac{\partial \Sigma(k',\omega')}{\partial k'_\mu}\right),
\label{pGpk}
\end{equation}
we obtain
\fulltext
\begin{equation}
\frac{\partial \Sigma(k,\omega)}{\partial k_\mu}
=-\mbox i
\int\frac{\mbox d^4 k'}{(2\pi)^4}
%\sum_{k'}
\Gamma^{(0)}(k,k')
\left\{G(k')^2\right\}^k%\infty
\left(
v_{k'}+\frac{\partial \Sigma(k',\omega')}{\partial k'}\right).
\label{delSigdelk'}
\end{equation}
\halftext
Note that the $k$-limit $\left\{G(k')^2\right\}^k$
appears in eq.~(\ref{pGpk}) 
when the derivative is taken with respect to $k'$
for fixed $\omega'$.
Comparing eq.~(\ref{delSigdelk'}) with 
\begin{equation}
\Lambda^k%\infty
_\mu(k')
=v_{k'\mu}-\mbox i
\int\frac{\mbox d^4 k''}{(2\pi)^4}
%\sum_{k'}
\Gamma^{(0)}(k',k'')
\left\{G(k'')^2\right\}^k%\infty
\Lambda^k%\infty
_\mu(k''),
\end{equation}
which is derived from eq.~(\ref{Lambda}),
we find
\begin{equation}
\Lambda^k%\infty
_\mu(k')=
v_{k'\mu}+\frac{\partial \Sigma(k',\omega')}{\partial k'_{\mu}},
\label{Lambdainf}
\end{equation}
hence,
\begin{equation}
v^*_{k'\mu}=z_{k'} \Lambda^k%\infty
_\mu(k'),
\label{Wardinf}
\end{equation}
because of eq.~(\ref{v*}).

Using the above results,
we can write the $k$-limit of $K_{\mu\nu}(k)$,
eq.~(\ref{Kmunu}), as
\begin{eqnarray}
K_{\mu\nu}^k%\infty
&=&-\frac{\mbox{i}}{\Omega}
%\frac{1}{\Omega}\sum_{k'}
\int\frac{\mbox d^4 k'}{(2\pi)^4}
v_{k'\mu}\left\{G(k')^2\right\}^k%\infty
\Lambda^k%\infty
_\nu(k')\\
&=&-\frac{\mbox{i}}{\Omega}
\int\frac{\mbox d^4 k'}{(2\pi)^4}
%\frac{1}{\Omega}\sum_{k'}
v_{k'\mu}
\frac{\partial}{\partial k'_\nu}G(k')
\label{Kinf0}\\
&=&\frac{\mbox{i}}{\Omega}
\int\frac{\mbox d^4 k'}{(2\pi)^4}
%\frac{1}{\Omega}\sum_{p'}
\frac{\partial v_{k'\mu}}{\partial k'_\nu}G(k'),
\label{Kinf}
\end{eqnarray}
where we used eqs.~(\ref{pGpk}) and (\ref{Lambdainf})
in the second line.
Substituting eq.~(\ref{Kinf}) for
the second term in the parenthesis of eq.~(\ref{sigmaKF}),
we can write the conductivity as
\begin{equation}
\sigma_{\mu\nu}(\omega)=\frac{\mbox{i}e^2}{\omega+\mbox{i}0}
\bigl(
K_{\mu\nu}(\omega+\mbox{i}0)-K_{\mu\nu}^k%\infty
\bigr).
\end{equation}
Thus,
the effective mass $m'$ in the Drude weight,
\begin{equation}
\mbox{Re}\sigma_{\mu\nu}(\omega)=
\pi e^2\left(\frac{n}{m'}\right)_{\mu\nu}
\delta(\omega) +\sigma_{\mbox{inc}},
\label{Drude+}
\end{equation}
is given by the difference between
the $\omega$-limit and the $k$-limit of $K_{\mu\nu}(k)$;
\begin{equation}
\left(\frac{n}{m'}\right)_{\mu\nu}
=%\mbox{Im}
%-\mbox i
%\bigl(
K_{\mu\nu}^\omega%0
-K_{\mu\nu}^k.%\infty
%\bigr).
\label{nm'K}
\end{equation}
%Here we note that
%$K_{\mu\nu}^k=\mbox{i}\langle
%\hat{\varepsilon}''_{\mu\nu}\rangle$ is an imaginary number.
%%Here we note that
%%$K_{\mu\nu}^k=\langle
%%\hat{\varepsilon}''_{\mu\nu}\rangle$ is a real number.
In other words, 
the Drude weight is determined by 
the singular (coherent) part of $G^2(k')$, eq.~(\ref{difG2}).
To estimate eq.~(\ref{nm'K}),
from eq.~(\ref{Lambda}) we obtain
\fulltext
\begin{equation}
\Lambda^\omega%0
_{\mu}(k) =\Lambda^k_{\mu}(k)
-\mbox i \int\frac{\mbox d^4 k'}{(2\pi)^4}
\Gamma^\omega%0
(k,k')
  \left(
    \left\{G(k')^2\right\}^\omega%0
    -\left\{G(k')^2\right\}^k%\infty
  \right)
\Lambda_\mu^k%\infty
(k').
\label{Lam0inf}
\end{equation}
%for which eq.~(\ref{G2r}) gives
%\begin{equation}
%\left\{G(k')^2\right\}^\omega%0
%-\left\{G(k')^2\right\}^k%\infty
%=2\pi \mbox i z_{k'}^2
%\delta(\mu-\varepsilon^*_{k'})
%\delta(\mu-\omega').
%\label{difG2}
%\end{equation}
Thus, eqs.~(\ref{difG2}), (\ref{Wardinf}), (\ref{Lam0inf})
lead to
\begin{eqnarray}
z_k%\mbox{Re}
\Lambda_\mu^\omega(k)=
v_{k\mu}^*+\sum_{k'}f(k,k')v_{k'\mu}^*\delta(\mu-\varepsilon^*_{k'})
\equiv j^*_{k\mu},
\label{Ward0}
\end{eqnarray}
\fulltext
where
\begin{equation}
f(k,k')\equiv z_k z_{k'}%\mbox{Re}
\Gamma^\omega(k,k').
\label{fkk'}
\end{equation}
Since we have
\begin{equation}
K^\omega%0
_{\mu\nu} =K^k%\infty
_{\mu\nu}
+\frac{1}{\Omega}
\int\frac{\mbox d^4 k'}{(2\pi)^4}
%\sum_p
\Lambda_\mu^k%\infty
(k')
  \left(
    \left\{G(k')^2\right\}^\omega%0
    -\left\{G(k')^2\right\}^k%\infty
  \right)
\Lambda_\mu^\omega%0
(k'),
\label{K0inf}
\end{equation}
\halftext
which is derived from eq.~(\ref{Kmunu}),
using eqs.~(\ref{Wardinf}), (\ref{difG2}) and  (\ref{Ward0}),
for eq.~(\ref{nm'K}) we finally obtain
\begin{equation}
\left(\frac{n}{m'}\right)_{\mu\nu}=
\frac{1}{\Omega}
\sum_k v^*_{k\mu} j^*_{k\nu} \delta(\mu-\varepsilon^*_k),
\label{nm'FLII}
\end{equation}
in agreement with eq.~(\ref{nm'FL}).
Hence, following the argument given below eq.~(\ref{nm'FL})
and using $j^*_{k}=v_k=k/m$,
we can conclude 
the absence of renormalization in the Drude weight
in Galilean invariant systems.
In particular,
when $\partial v_{k\mu}/\partial k_\nu=\delta_{\mu\nu}/m$,
eq.~(\ref{Kinf}) gives
\begin{eqnarray}
%\mbox i
K_{\mu\nu}^k%\infty
&=&
\frac{\delta_{\mu\nu}}{m}
\frac{\mbox i}{\Omega}
\int\frac{\mbox d^4 k'}{(2\pi)^4}G(k')
=-\frac{n}{m}\delta_{\mu\nu}.
\end{eqnarray}
Therefore, in this case, the sum rule,
\begin{eqnarray}
\int^\infty_{-\infty}\mbox{Re}\sigma_{\mu\nu}(\omega)
\mbox d \omega&=&
%\frac{\pi e^2}{\Omega}\sum_k
%\left\langle \frac{\partial^2 \varepsilon_k}{\partial k_\mu\partial
%k_\nu}\right\rangle.
\frac{\pi e^2}{\Omega}\left\langle
\hat{\varepsilon}''_{\mu\nu}
\right\rangle
\label{sumrule}\\
&=&\frac{\pi n e^2}{m}\delta_{\mu\nu},
\end{eqnarray}
is saturated with the Drude weight, hence $K_{\mu\nu}^\omega=0$.
In general, a lost weight in the coherent part is transfered
to an incoherent part;
\begin{equation}
\left(\frac{n}{m}\right)_{\mu\nu}
-\left(\frac{n}{m'}\right)_{\mu\nu}=
-\frac{1}{\pi} \int_{-\infty}^{\infty}
\frac{\mbox{Im}K_{\mu\nu}(\omega')}{\omega'}
\mbox{d}\omega'\ge 0,\label{sumrule2}
\end{equation}where
we defined the effective mass in the total weight,
\begin{equation}
\left(\frac{n}{m}\right)_{\mu\nu}\equiv
\frac{1}{\Omega}\left\langle
\hat{\varepsilon}''_{\mu\nu}
\right\rangle.
\label{nm}
\end{equation}
%The right-hand side of eq.~(\ref{sumrule2})
%cannot be described within the framework of the Fermi liquid theory.
The incoherent part of the spectrum cannot be described 
in the framework of the Fermi liquid theory.~\cite{rf:com}

In the same manner as we did for $G(k)$,
we can derive the Ward identity for 
the Green's function $G_p(k)$ defined in eq.~(\ref{Gpp'}).
Let us introduce $\varepsilon^*_{kp}$ defined by
\begin{equation}
%\mbox{Re}G_p^{-1}(k,\varepsilon^*_{kp})=
\varepsilon^*_{kp}-
\varepsilon_{k+p}-\Sigma_p(k,\varepsilon^*_{kp}-\mu)=0.
\label{G-1p=0}
\end{equation}
Differentiating this with respect to $p$,
we find
\begin{equation}
\left.\frac{\partial \varepsilon^*_{kp}}{\partial p}\right|_{p=0}
=z_k\left(
v_{k}+\left.\frac{\partial }{\partial p}
  \Sigma_p(k,0)
\right|_{p=0}
\right).
\label{v*p}
\end{equation}
%BBBBBBBBBBBBBBBBBBBBBBBBBBBB
%\fulltext
%\begin{equation}
%\Lambda^\omega%0
%_{\mu}(p) =\Lambda^k%\infty
%_{\mu}
%-\mbox i \int\frac{\mbox d^4 k'}{(2\pi)^4}
%\Gamma^k
%(p,k')
%  \left(
%    \left\{G(k')^2\right\}^\omega%0
%    -\left\{G(k')^2\right\}^k%\infty
%  \right)
%\Lambda_\mu^\omega
%(k').
%\label{Lam0inf2}
%\end{equation}
%\begin{eqnarray}
%z_k\mbox{Re}\Lambda_\mu^\omega(k)=
%j_{k\mu}^*=v^*_{k\mu}
%+\sum_{k'}A(k,k')j_{k'\mu}^*\delta(\mu-\varepsilon^*_{k'})
%\end{eqnarray}
%\begin{equation}
%A(k,k')\equiv z_k z_{k'}\Gamma^k(k,k').\qquad(\mbox{real })
%\end{equation}
%CCCCCCCCCCCCCCCCCCCCCCCCCCCCCCCCCC
Similarly as in eqs.~(\ref{pSpk'}) - (\ref{Wardinf}),
using
\begin{equation}
\frac{\partial \Sigma_p(k',\omega')}{\partial p_\mu}
=-\mbox i
\int\frac{\mbox d^4 k''}{(2\pi)^4}
\Gamma^{(0)}(k',k'')\frac{\partial}{\partial p_\mu}G_p(k'',\omega''),
\end{equation}
and
\begin{equation}
\frac{\partial}{\partial p_\mu}G_p(k',\omega')=
\left\{G(k')^2\right\}^\omega
\left(
v_{k'\mu}+\frac{\partial \Sigma_p(k',\omega')}{\partial p_\mu}\right),
\label{pGpkp}
\end{equation}
we get
\fulltext
\begin{equation}
\frac{\partial \Sigma_p(k',\omega')}{\partial p_\mu}
=-\mbox i
\int\frac{\mbox d^4 k''}{(2\pi)^4}
%\sum_{k'}
\Gamma^{(0)}(k',k'')
\left\{G(k'')^2\right\}^\omega%0
\left(
v_{k''\mu}+\frac{\partial \Sigma_p(k'',\omega'')}{\partial
p_\mu}\right).
\end{equation}
\halftext
Here the regular function $\left\{G(k'')^2\right\}^\omega$
in the $\omega$-limit should be used.
This is because
%related to the fact that 
the pole $\varepsilon^*_{kp}$ of $G_p(k,\omega)$
adiabatically moves to $\varepsilon^*_{k}$ as a function of $p$
without crossing the Fermi surface, 
since
the Fermi surface for $G_p(k')$ is held fixed as a function of $p$.
Thus, because of 
\begin{equation}
\Lambda^\omega_\mu(k')
=v_{k'\mu}-\mbox i
\int\frac{\mbox d^4 k''}{(2\pi)^4}
%\sum_{k'}
\Gamma^{(0)}(k',k'')
\left\{G(k'')^2\right\}^\omega
\Lambda^\omega_\mu(k''),
\end{equation}
which is obtained from eq.~(\ref{Lambda}),
we find
\begin{equation}
\Lambda^\omega_\mu(k')=
v_{k'\mu}+\frac{\partial \Sigma_p(k',\omega')}{\partial p_\mu},
\end{equation}
and eq.~(\ref{v*p}) gives
\begin{equation}
\left.\frac{\partial \varepsilon^*_{kp}}{\partial p}\right|_{p=0}
%=z_{k'} \Lambda^\omega_\mu(k')
=j^*_{k\mu},
\label{Wardinfp}
\end{equation}
where we used eqs.~(\ref{Ward0}) and (\ref{v*p}).
Equation (\ref{Wardinfp}) is to be compared with 
eq.~(\ref{QPv}).
%, and also with eq.~(\ref{Wardinf}).
In particular, under the assumption 
\begin{equation}
 v_{k}+v_{k'}-v_{k+q}-v_{k'-q} = 0,
\label{totvne00}
\end{equation}
we have
\begin{equation}
v_k\frac{\partial \Sigma_p(k,\omega)}{\partial \omega}
+\frac{\partial \Sigma_p(k,\omega)}{\partial p}=0.
\end{equation}
It is easy to check this equation 
for a few diagrams of lower order.
Therefore, in this case, 
we see from eqs.~(\ref{v*p}) and (\ref{Wardinfp}) that
the quasiparticle current
is not renormalized; $j^*_k=v_k$. 
This equation led us to 
conclude the absence of renormalization in $D$.

To derive eq.~(\ref{J=delEdelp}),
we write the $\omega$-limit of $K_{\mu\nu}(k)$ as
\begin{eqnarray}
K_{\mu\nu}^\omega&=&-\frac{\mbox{i}}{\Omega}
\int\frac{\mbox d^4 k'}{(2\pi)^4}
v_{k'\mu}\left\{G(k')^2\right\}^\omega
\Lambda^\omega_\nu(k')\nonumber\\
&=&-\frac{\mbox{i}}{\Omega}
\int\frac{\mbox d^4 k'}{(2\pi)^4}
v_{k'\mu}
\frac{\partial}{\partial p_\nu}G_p(k'),
\label{K0}
\end{eqnarray}
which corresponds to eq.~(\ref{Kinf0}) for $K_{\mu\nu}^\infty$.
Subtracting eq.~(\ref{Kinf}) from eq.~(\ref{K0}),
\begin{eqnarray}
K_{\mu\nu}^\omega-K_{\mu\nu}^k
&=&-\frac{\mbox{i}}{\Omega}
\int\frac{\mbox d^4 k'}{(2\pi)^4}
\left(
v_{k'\mu}
\frac{\partial}{\partial p_\nu}G_p(k')
+
\frac{\partial v_{k'\mu}}{\partial k'_\nu}G(k')
\right)\nonumber\\
&=&-\frac{\mbox{i}}{\Omega}
\int\frac{\mbox d^4 k'}{(2\pi)^4}
\left.\frac{\partial}{\partial p_\nu}
\bigl(v_{k'+p\mu}G_p(k')\bigr)\right|_{p=0}.
\nonumber
\end{eqnarray}
Therefore, in terms of eqs.~(\ref{Jp}) and (\ref{Jp=delEp}), 
for eq.~(\ref{nm'K})
we finally obtain
\begin{eqnarray}
\left(\frac{n}{m'}\right)_{\mu\nu}
&=&
\frac{1}{\Omega}\frac{\partial J_{p\mu}}{\partial p_\nu}
=\frac{1}{\Omega}\frac{\partial^2 E_p}{\partial p_\mu\partial p_\nu}.
\label{J=delEdelpII}
\end{eqnarray}

%%%%%%%%%%%%%%%%%%%%%%%%%%
%%%%%%%%%%%%%%%%%%%%%%%%%%
%%%%%%%%%%%%%%%%%%%%%%%%%%
%%%%%%%%%%%%%%%%%%%%%%%%%%
%%%%%%%%%%%%%%%%%%%%%%%%%%
%%%%%%%%%%%%%%%%%%%%%%%%%%
%%%%%%%%%%%%%%%%%%%%%%%%%%
\section{Perturbation Theory}
\label{PT}
\subsection{Linear response theory}
The effective mass in the Drude weight,
defined in eq.~(\ref{nm'FL}) or eq.~(\ref{nm'FLII}),
which is derived on the basis of the Fermi liquid theory,
gives us a definite picture on
how the Drude weight is renormalized by the many-body effect.
However, to estimate the effective mass perturbatively,
the formula (\ref{J=delEdelpII}) is practically useful.
Not only to verify this formula
but to see how Umklapp processes affect the Drude weight,
we apply a finite-order perturbation theory in the following.

In the spectral representation,
the dynamical conductivity (\ref{sigmaKF})
is written as
\fulltext
\begin{equation}
\sigma_{\mu\nu}(\omega)=
\frac{e^2}{\mbox i\omega}
\biggl(\frac{\mbox i}{\Omega}
\int^\infty_{-\infty}
\langle 0|\mbox T \hat v_\mu(t)\hat v_\nu |0\rangle
\exp(\mbox i\omega t)\mbox d t
-\frac{1}{\Omega}\langle 0|
\hat{\varepsilon}''_{\mu\nu}
|0 \rangle \biggr).
\label{dyncon}
\end{equation}
\halftext
For $\hat v_\mu$ and \(\hat{\varepsilon}''_{\mu\nu}\)
in this expression, we use eqs.~(\ref{hatv}) and (\ref{hatep''}),
where the sum over the spin component $\sigma$ is implicitly 
assumed in the summation $\Sigma_k$.
Then, taking the limit $\omega\rightarrow 0$,
we get
%~\cite{rf:Kohn,rf:SS}
%eq.~(\ref{conductivity})
%the Drude weight in terms of 
\begin{equation}
\left(\frac{n}{m'}\right)_{\mu\nu}
\equiv
-\frac{2}{\Omega}\sum_{r\ne 0}
\frac{\langle 0|\hat{v}_\mu|r\rangle
\langle r|\hat{v}_\nu|0\rangle}{E_r-E_0}
+\frac{1}{\Omega}\langle 0|
%\hat{\frac{\partial^2\varepsilon_k}{\partial k_\mu\partial k_\nu}}
\hat{\varepsilon}''_{\mu\nu}
|0 \rangle.
\label{nm'sp}
\end{equation}
%which is the spectral representation of eq.~(\ref{nm'withG}).
In this expression, the sum is taken over the exact excited states
$|r\rangle$ of the system.
%In the same manner as we discussed below eq.~(\ref{delmup}),
The right-hand side  of eq.~(\ref{nm'sp})
is derived also from the energy shift $E_p-E_0$ caused by 
the perturbation,
\begin{eqnarray}
H_p-H&\equiv&\sum_{k\sigma} (\varepsilon_{k+p}-\varepsilon_k)
 \hat{c}^\dagger_{k\sigma}\hat{c}_{k\sigma}\nonumber\\
&\simeq&
{\mib p}\mib \cdot\hat{\mib v}+
\frac{1}{2}\sum_{\mu,\nu}p_\mu p_\nu
\hat{\varepsilon}''_{\mu\nu}.
\label{Hp-H}
\end{eqnarray}
As a result we obtain the relation
\begin{equation}
\frac{\partial^2 E_p}{\partial p_\mu\partial p_\nu}
= \Omega\left(\frac{n}{m'}\right)_{\mu\nu},
\end{equation}
for 
\begin{equation}
E_p=\langle p|H|p\rangle=\langle 0|H_p|0\rangle,
\label{Ep}
\end{equation}
where $|p\rangle$ is the ground state of $H_p$.

At this point,  it is clear that 
the method given here has a close connection with
Kohn's discussion~\cite{rf:Kohn,rf:SS} that
the effective mass defined in eq.~(\ref{Drude})
is derived by investigating
variation of the ground state energy under changes in 
the boundary condition of a {\it finite} system.
In a finite system the change of the boundary condition causes
a uniform shift of discrete momentum quantum numbers.
%while we are treating an infinitely large system from the outset.
The point is that, in our theory,
the chemical-potential shift eq.~(\ref{Hp-H})
should be treated as a perturbation.
Since the perturbation adiabatically applied to the system 
cannot deform the Fermi surface,
we cannot reach $|p\rangle$ from $|0\rangle$
by perturbation theory,
but we are led to the excited state
of the energy $E_p$, eq.~(\ref{Ep}),
instead of the ground state of $H_p$.
In practice, 
if $E_0$ ($=\langle 0|H|0\rangle$)
is expressed in terms of bare quantities and 
interaction parameters, $E_p$ is easily obtained from $E_0$
by replacing the distribution function $n_k$ with $n_{k-p}$. 
%with the other quantities, such as $\varepsilon_k$, 
%being held fixed.
Thus, as a method to calculate the Drude weight, our theory 
is equivalent to Kohn's argument, so that
%as an equation to estimate the Drude weight 
%of an arbitrary state $|0\rangle$, 
eq.~(\ref{J=delEdelpII}) is generally applied with
 the proviso that $E_p$ is
regarded as the energy of the state $|p\rangle$ 
derived from $|0\rangle$.
The investigation made in the preceding sections
is still instructive, since 
it clarifies the physical meaning of the Drude weight $D$:
We must assume $D>0$ for 
the Fermi liquid state to be stable against
the shift of the Fermi surface in the momentum space.

%against the perturbation
%to displace the Fermi surface by $p$ in the momentum space.

%(However, it is noted that $|p\rangle$ is not reached
%adiabatically from $|0\rangle$
%by the perturbation theory.)

\subsection{Drude weight}
To evaluate eq.~(\ref{nm'sp}) by perturbation theory,
the ground state $|0\rangle$ is expanded
up to terms of the second order in
the interaction
\(V'\equiv V-\langle V\rangle,\)~\cite{rf:LL}
\begin{eqnarray}
|0\rangle &=&|0\rangle_0+|0\rangle_1+|0\rangle_2,\nonumber\\
|0\rangle_1&=&-\sum_{r\ne 0}\frac{|r\rangle\langle
r|V'|0\rangle_0}{E_r-E_0},\\
|0\rangle_2&=&\sum_{r\ne 0,s\ne 0}
\frac{|r\rangle\langle r|V'|s\rangle\langle s|V'|0\rangle_0}
{(E_r-E_0)(E_s-E_0)}\nonumber\\
&&-\frac{|0\rangle_0}{2}\sum_{r\ne 0}\frac{\left|\langle
r|V'|0\rangle_0\right|^2}{(E_r-E_0)^2}.
\end{eqnarray}
To put it concretely, let us study the Hubbard model
\begin{equation}
H=T+V=
\sum_{k,\sigma}\varepsilon_k \hat{c}^\dagger_{k\sigma}
\hat{c}_{k\sigma}
+\sum_i U\hat{n}_{i\uparrow}\hat{n}_{i\downarrow}.
\label{Hubbard}
\end{equation}
%in a square lattice $\varepsilon_k=-2t(\cos(k_x)+\cos(k_y))$.
We assume that the effective mass 
in eq.~(\ref{nm'sp}) becomes isotropic in the $x$-$y$ plane,
$m'_{\mu\nu}\equiv m'\delta_{\mu\nu}$.
In the zeroth order in $U$,
the second term of eq.~(\ref{nm'sp}) gives
\begin{eqnarray}
\left(\frac{n}{m'}\right)_0&=&
\,_0\langle 0|
\hat{\varepsilon}''_{\mu\mu}
|0 \rangle_0\\
&=&\frac{1}{\Omega}\sum_{k\sigma} n_{k}
 \frac{\partial^2 \varepsilon_k}
{\partial k_\mu^2}\label{1/m0}\\
&=&\frac{1}{\Omega}
\left.\frac{\partial^2}{\partial p^2} E_p^{(0)}\right|_{p=0},
\label{nm'=EI}
\end{eqnarray}
where
\begin{equation}
n_{k}=\,_0\langle 0|
 \hat{c}^\dagger_{k\sigma}\hat{c}_{k\sigma}
|0 \rangle_0,
\end{equation}
and
\begin{equation}
E_p^{(0)}=
\sum_{k\sigma} n_{k} \varepsilon_{k+p}
=
\sum_{k\sigma} n_{k-p} \varepsilon_{k}
\label{E0},
\end{equation}
or,  similarly we have
\begin{equation}
\left(\frac{n}{m'}\right)_0=-
\frac{1}{\Omega}\sum_{k\sigma}
\frac{\partial n_{k}}{\partial k_\mu}
\frac{\partial \varepsilon_k}{\partial k_\mu}
=-\frac{1}{\Omega}\sum_{k\sigma}
\frac{\partial n_{k}}{\partial\varepsilon_k}
v_{k\mu}^{2}.
\label{1/m02}
%\\&=&\frac{1}{\Omega}\sum_p n_p \frac{\partial^2 \varepsilon_p}
%{\partial p_\mu^2}.
%\label{1/m0}
\end{equation}
%In these expressions $\sim$(\ref{1/m02}),
%instead of eq.~(\ref{1/m0}).
It is noted in eq.~(\ref{E0}) that
shifting the distribution $n_{k}\rightarrow n_{k- p}$
is formally equivalent to
replacing $\varepsilon_{k+p}$ by $\varepsilon_{k}$
for fixed $n_{k}$.
Accordingly, in eq.~(\ref{1/m0}),
$(n/m')_0$ is expressed 
as an average of $\partial^2 \varepsilon_k/\partial k_\mu^2$
over the states $\it below$ the Fermi level,
while in eq.~(\ref{1/m02}) it is given by
an average of $v_{k\mu}^2$ over the states $\it at$ the Fermi level,
as in eq.~(\ref{nm'FLII}).
To estimate $1/m'_0$ numerically,
eq.~(\ref{1/m0}) is more suitable than eq.~(\ref{1/m02}).

For the correction $(n/m')_2$ of order $O(U^2)$,
from the first term of eq.~(\ref{nm'sp}) we obtain 
\fulltext
\begin{equation}
-\frac{2}{\Omega}\sum_{r\ne 0}
\frac{|\langle r|\hat{v}_x|0\rangle_1|^2}{E_r-E_0}
=
-2\frac{U^2}{\Omega^3}\sum_{k_1\sim k_4}\frac{
(1-n_{k_1})(1-n_{k_2})n_{k_3}n_{k_4}}
{(\varepsilon_{k_1}+\varepsilon_{k_2}
-\varepsilon_{k_3}-\varepsilon_{k_4})^3}
(v_{k_1x}+v_{k_2x}-v_{k_3x}-v_{k_4x})^2\delta_{k_1+k_2-k_3-k_4},
\label{nm'2I}
\end{equation}
where
\begin{equation}
\delta_{q}\equiv\frac{1}{\Omega}\sum_i \mbox e^{-\mbox iqr_i},
\end{equation}
and from the second term of eq.~(\ref{nm'sp}),
\begin{eqnarray}
\frac{1}{\Omega}
\lefteqn{\left(_1\langle 0|
\hat{\varepsilon}''_{\mu\mu}
|0 \rangle_1+
\, _0\langle 0|
\hat{\varepsilon}''_{\mu\mu}
|0 \rangle_2+
\, _2\langle 0|
\hat{\varepsilon}''_{\mu\mu}
|0 \rangle_0\right)}
\nonumber\\
&&=
\frac{U^2}{\Omega^3}\sum_{k_1\sim k_4}\frac{
(1-n_{k_1})(1-n_{k_2})n_{k_3}n_{k_4}}
{(\varepsilon_{k_1}+\varepsilon_{k_2}
-\varepsilon_{k_3}-\varepsilon_{k_4})^2}
\left(\frac{\partial^2\varepsilon_{k_1}}{\partial k_x^2}
+\frac{\partial^2\varepsilon_{k_2}}{\partial k_x^2}
-\frac{\partial^2\varepsilon_{k_3}}{\partial k_x^2}
-\frac{\partial^2\varepsilon_{k_4}}{\partial k_x^2}\right)
\delta_{k_1+k_2-k_3-k_4}.
\label{nm'2II}
\end{eqnarray}
As a result,
\begin{equation}
\protect
\left(\frac{n}{m'}\right)_2
=\mbox{eq.~(\ref{nm'2I})}+\mbox{eq.~(\ref{nm'2II})},
\label{nm'2}
\end{equation}
%summing eqs.~(\ref{nm'2I}) and (\ref{nm'2II}),
and we find
\begin{equation}
\left(\frac{n}{m'}\right)_2
=\frac{1}{\Omega}
\left.\frac{\partial^2}{\partial p^2} E_p^{(2)}\right|_{p=0},
\label{nm'=EII}
\end{equation}
where
\begin{eqnarray}
E_p^{(2)}&=&
-\frac{U^2}{\Omega^2}\sum_{k_1\sim k_4}\frac{
(1-n_{k_1})(1-n_{k_2})n_{k_3}n_{k_4}}
{\varepsilon_{k_1+p}+\varepsilon_{k_2+p}
-\varepsilon_{k_3+p}-\varepsilon_{k_4+p}}
\delta_{k_1+k_2-k_3-k_4}\\
&=&
-\frac{U^2}{\Omega^2}\sum_{k_1\sim k_4}\frac{
(1-n_{k_1-p})(1-n_{k_2-p})n_{k_3-p}n_{k_4-p}}
{\varepsilon_{k_1}+\varepsilon_{k_2}
-\varepsilon_{k_3}-\varepsilon_{k_4}}
\delta_{k_1+k_2-k_3-k_4}.
\end{eqnarray}
\halftext

The above results,
following eqs.~(\ref{nm'=EI}) and (\ref{nm'=EII}),
verify the relation (\ref{J=delEdelpII})
up to terms of the second order in $U$.
In particular, eqs.~(\ref{nm'2I}) and (\ref{nm'2II}) indicate
$(n/m')_{2}=0$ for  
\begin{equation}
v_{k}+v_{k'}-v_{k+q}-v_{k'-q}= 0,
\label{totv0}
\end{equation}
i.e., the violation of the group-velocity conservation 
leads to renormalization of the Drude weight,
in accordance with the note made below eq.~(\ref{totvne00}).
Recently this point was discussed
by Maebashi and Fukuyama.~\cite{rf:MF}\

In Fig.~\ref{fig}, the ratio $n/m'$
calculated with eqs.~(\ref{nm'=EI}) and (\ref{nm'=EII})
are displayed as a function of $n$ for $U=0$ and $U=1$ 
for the square lattice with $4t=1$ and also for 
the case where the next-nearest-neighbor hopping $t'=-0.2t$
is included.
%in a square lattice $\varepsilon_k=-2t(\cos(k_x)+\cos(k_y))$.
In the figure, $n/m$ for the total weight defined in
eq.~(\ref{nm}), which is given by 
eqs.~(\ref{1/m02}) and (\ref{nm'2II}), 
is also shown.
It is clear from the figure that
the many-body effect modifies the Drude weight,
\[D=\frac{\pi e^2n}{m'}.\]
For $U=1$, we see $n/m'$ decreases precipitously
as $n$ approaches the van Hove singularity.
%half filling, $n=1$.
In the second-order correction $-(n/m')_{2}$, 
we found that the part $-(n/m')_{2}^{\rm n}$ 
that is not due to Umklapp processes 
varies monotonically.
For example,
in %the left of
Fig.~\ref{fig} (a),
$-(n/m')_{2}^{\rm n}$ increases almost linearly 
up to less than $0.03$ as $n$ approaches half filling.
The small and non-singular behavior of $-(n/m')_{2}^{\rm n}$
due to normal processes is understood by observing that 
the relevant normal processes 
are not restricted in a narrow
%do not lie in the low-energy 
region around the Fermi level,
where the effect of the van Hove singularity is prominent:
For the Fermi surfaces of our example, energy conservation excludes 
normal processes at the Fermi level.~\cite{rf:MF}\
Thus in our example of a single-band model,
the renormalization of the Drude weight and 
the conspicuous behavior shown in the figure 
are primarily caused by Umklapp processes, that is, 
the renormalization due to the normal processes violating 
the velocity conservation~\cite{rf:MF}
are qualitatively unimportant here.

\begin{fullfigure}[t]
\centerline{
\epsfile{file=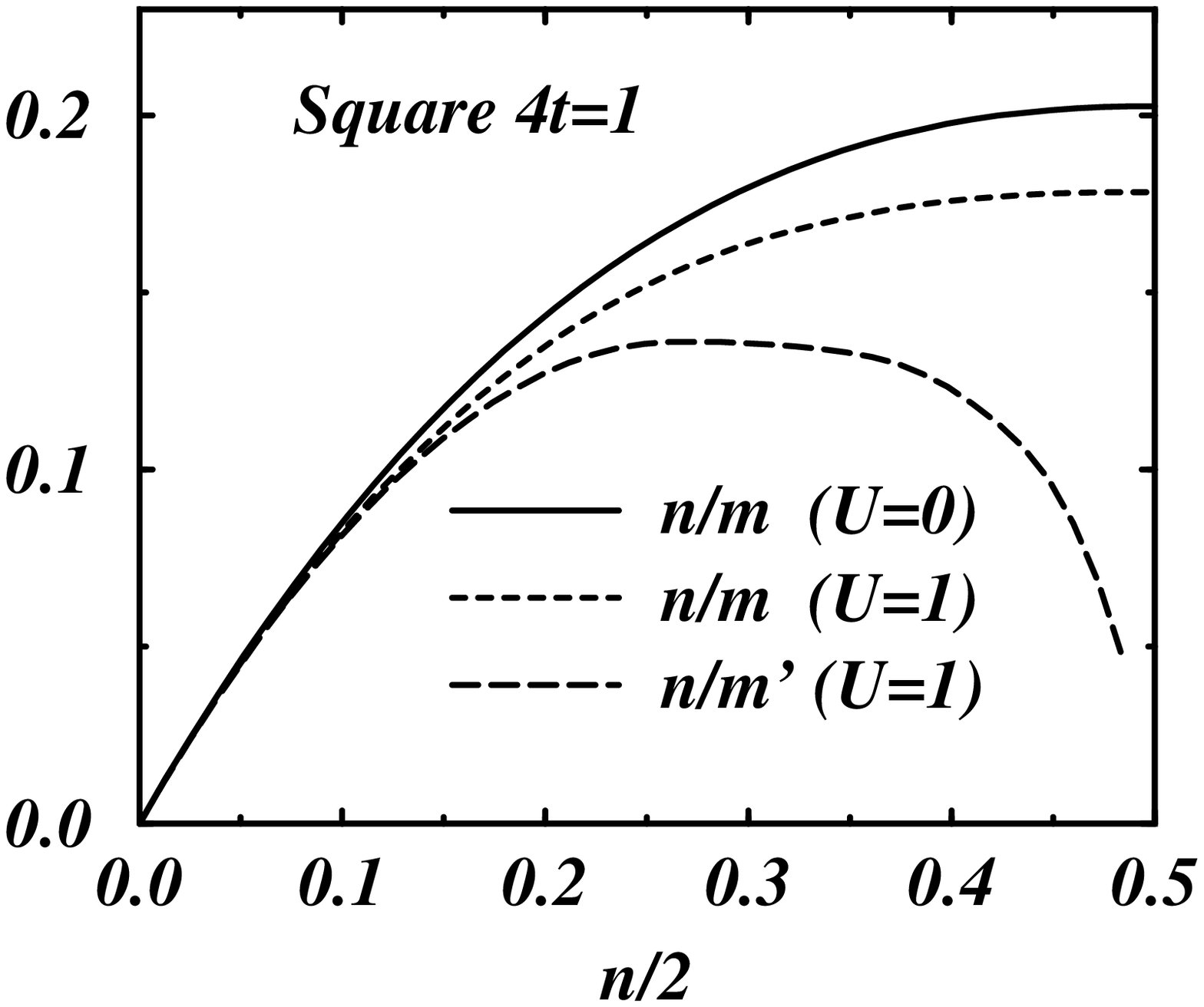,width=6.6cm}
%}
% \hspace{6mm}
%\centerline{
\epsfile{file=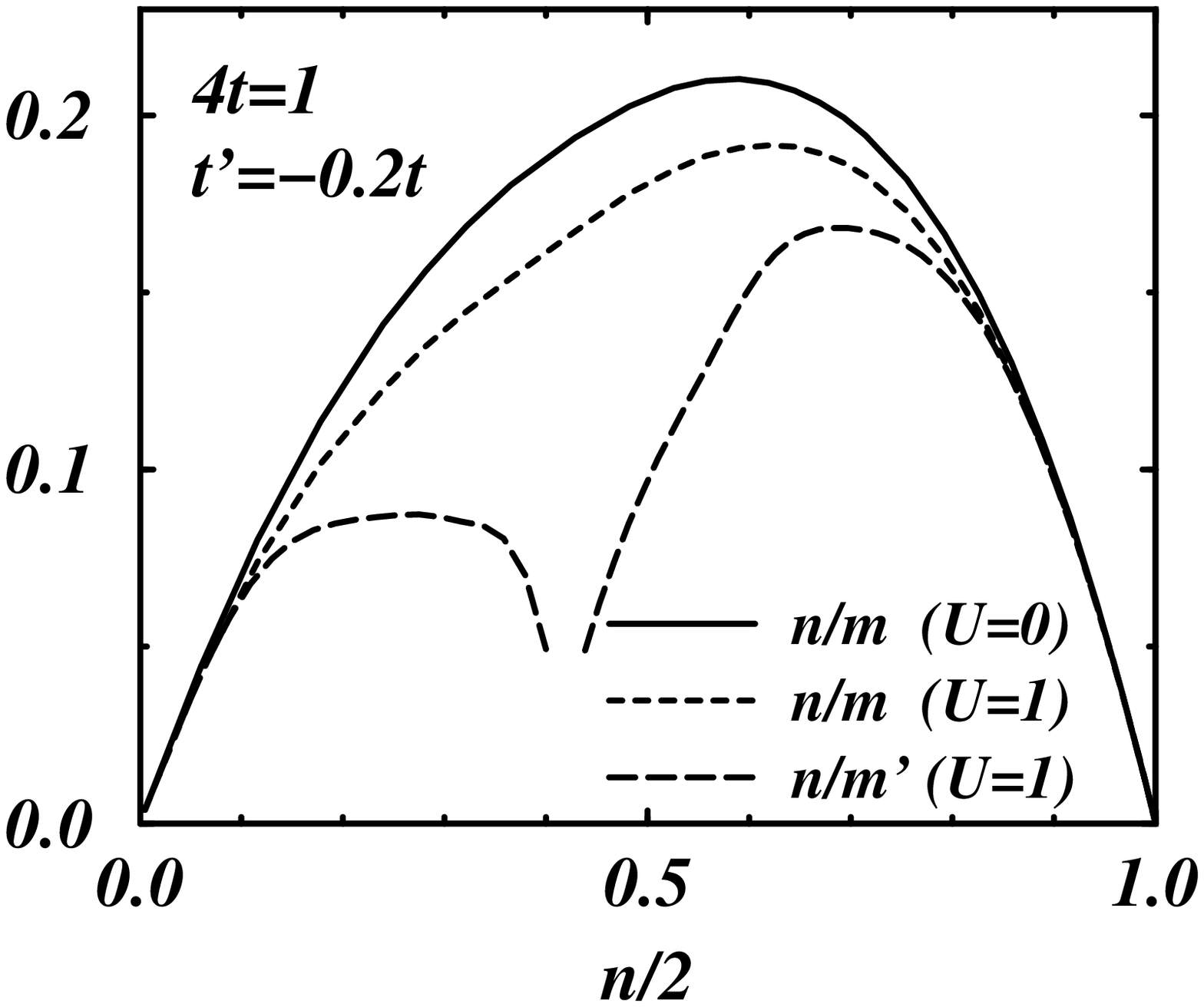,width=6.6cm}
}
\caption{(a) 
$n/m' $ for $U=0$, $U=1$ and $n/m$ for $U=1$ 
are displayed as a function of $n/2$ 
for the Hubbard model in the square lattice with $4t=1$ (left).
(b) The same quantities in the presence of
the next-nearest-neighbor hopping $t'=-0.2t$ (right).
The Drude weight $D$ and the total weight of $\sigma(\omega)$ 
are proportional to $n/m'$ and $n/m$, respectively.
The curves for $n/m'$ ($U=1$) 
are not shown around the van Hove singularity.
}
\label{fig}
\end{fullfigure}

We have \({\mbox d}D/{\mbox d}n=\pi e^2/m'>0\)
for the Drude weight in an isotropic system,
since $m'$($\equiv m$) in this case is independent of $n$.
Correspondingly, even in a lattice system,
it is convenient to
regard current-carrying carriers as particles (holes)
when \({\mbox d}D/{\mbox d}n\) is positive (negative).
For instance, in the $U=0$ Hubbard model on a square lattice, 
we have \({\mbox d}D/{\mbox d}n>0\) for $n<1$
as shown by the solid curve in Fig.~\ref{fig},
while \({\mbox d}D/{\mbox d}n<0\) for $n>1$.
Thus for $n>1$  the total current 
may be well regarded as being carried by holes in the filled band.
On the other side, in one dimensional systems, $D$
vanishes continuously not only for $n\rightarrow 0$
and $n\rightarrow 2$ but also for $n\rightarrow 1$
for finite interaction $U> 0$.~\cite{rf:1d} 
Therefore, in this case, even in the underdoped regime $n\LA 1$
the current-carrying carrier
is regarded as holes doped in an insulating state at $n=1$.
From this point of view, it is interesting to note that
our result for $n/m'$ $(\propto D)$
shows that \({\mbox d}D/{\mbox d}n \) for  $U/4t=1$
first decreases from a positive value as $n$ increases from 0,
and finally it becomes negative for $n\LA 1$.
The result indicates a hole-like behavior around half filling,
$n\LA 1$.
This behavior is caused by the fact that the slope
\({\mbox d}D/{\mbox d}n\) for $U=0$ vanishes as $n\rightarrow 1$,
while a finite negative correction $(n/m')_2$ decreases in this limit, 
as Umklapp processes becomes effective.
When the Umklapp processes are effective, 
the Drude weight is reduced by the many-body effect,
so that a large coupling constant $\rho^* U$ gives rise to
a large difference $(n/m')_0-(n/m')_2$,
where $\rho^*$ is the density of states at the Fermi level.
Thus the decreasing 
derivative \({\mbox d}D/{\mbox d}n \) for $n\LA 1$ 
as a function of $U$ $(>0)$ is generally expected
even in the weak coupling regime.
For example, in our model, we assumed 
the coupling constant $U/W=0.5$ for the bandwidth $W=8t$.
It is noted however that 
our results do not imply $D\propto |1-n|$ for $n\rightarrow 1$
as in one dimensional systems, and that the hole-like behavior
determined from the sign of \({\mbox d}D/{\mbox d}n \)
has nothing to do with the shape of the Fermi surface, namely, 
whether it is closed around $%(k_x,k_y)
\mib{k}=(\pi,\pi)$.
%which does not appear explicitly in the above calculation.
In effect,
owing to the complete-nesting property occurring at $n=1$,
we should have to take account of antiferromagnetic ordering
which sets in around half filling.
%Accordingly, in a proper treatment, $D$ must vanish
%for $n\rightarrow 1$.
Our results shown in Fig.~\ref{fig}
are valid when the ground state is a paramagnetic Fermi liquid.

In a strong coupling regime,
the decrease in the Drude weight around half filling 
is reproduced by a numerical technique.~\cite{rf:tJ1}\
In our example of the weak coupling theory, however,
the steep decrease in $(n/m')_2$ of eq.~(\ref{nm'2})
is mainly caused by eq.~(\ref{nm'2I}), 
as is clear from the figure. % by comparing $n/m$ and $n/m'$.
In other words, the behavior $D\rightarrow 0$ as $n\rightarrow 1$ 
is ascribed to Umklapp processes becoming effective in this limit, 
rather than the decrease of the total weight $n/m$, 
eq.~(\ref{nm}).
%~\cite{rf:tJ1}\
Therefore, 
in regard to the mechanism to reduce $D$,
our result should not be simply compared with 
the situation in the strong coupling regime, 
as in the $t$-$J$ model,~\cite{rf:tJ}\
where the decreasing behavior of $D$
would be mainly due to 
$0\le D/\pi e^2= n/m'\le n/m\rightarrow 0$.

\section{Discussions}
\label{sec:D}
From the above results, it is remarked that
the large quasiparticle mass $m^*$ does not necessarily imply 
the enhanced mass $m'$ in the Drude weight.
This point is 
evident from the fact that, in an isotropic system,
the former is affected by the many-body effect while the latter is not.
Physically this means that
a small Drude weight does not necessarily suggest
a large specific heat coefficient.
Thus, although $m'$ and $m^*$ are related with each other
(eqs.~(\ref{1/m'ne1/m*}) and (\ref{jp})),
they are practically regarded as independent quantities.
In particular, this should be the case
when we know little about the function $f(k,k')$ 
by the first principle calculation.
We introduced the mass $m'$ for the sake of convenience
so that it represents the mass per particle of the system as a whole,
\[ \frac{E_p}{N}=\frac{p^2}{2m'}.\]
This is obtained from eq.~(\ref{J=delEdelpII}).
In this form it may become even clear why
electron-electron interactions do not modify $m'$
in a Galilean invariant system; where
the electron-electron interaction
cannot change the inertial mass of the system as a whole.
The effective mass $m'$ thus must be positive
for the system to be stable.~\cite{rf:negD}\
As is clear from eq.~(\ref{1/m'ne1/m*}),
to distinguish $m'$ from $m^*$,
we must distinguish between the current and velocity of quasiparticle,
and thereby we should not neglect the vertex correction.
The difference is due to the Fermi liquid effect.

If a Fermi liquid state were continuously driven to
the metal-insulator transition, 
we should have $m'\rightarrow \infty$. 
To see how this occurs,
from eqs.~(\ref{nm'FL}) and (\ref{jp}) 
 we may write
%\begin{equation}
%\frac{1}{m'}\simeq \frac{1}{m^*}\left(
%1+\frac{F_1}{3}\right),
%\end{equation}
%\begin{equation}
%\frac{F_1}{3}\equiv
%\frac{
%\sum_{k, k'}f(k,k')\mib{k\cdot k'}
%\delta(\mu-\varepsilon_{k})\delta(\mu-\varepsilon_{k'})}
%{\sum_{k}k^2\delta(\mu-\varepsilon_{k})}.
%\end{equation}
\begin{equation}
\frac{1}{m'}=\frac{1}{m^*}\left(1+\frac{F^s_1}{3}\right),
\end{equation}
where
\begin{equation}
\frac{n}{m^*}\equiv \frac{1}{3}
\sum_k \mib{v_k\cdot v_{k}}\delta(\mu-\varepsilon_k),
\end{equation}
\begin{equation}
\frac{F^s_1}{3}\equiv
\frac{\sum_{k,k'}f(k,k')
\mib{v_k\cdot v_{k'}}
\delta(\mu-\varepsilon_k)\delta(\mu-\varepsilon_{k'})}{
\sum_k \mib{v_k\cdot v_{k}}\delta(\mu-\varepsilon_k)}.
\end{equation}
Thus, to obtain $m'\rightarrow \infty$, 
we must have either (i) $m^*\rightarrow\infty$ or 
(ii) $F^s_1\rightarrow -3$.
The former, if the coupling $F^s_1$ remains unaffected,
corresponds to reduction of energy scale of the system,
with no qualitative change of the low-energy physics.
In this case, the total weight $n/m$, eq.~(\ref{nm}), 
will be reduced as well.
The latter, usually unnoticed, is a nontrivial possibility,
which reminds us of the ferromagnetic instability 
$F_0^a \rightarrow -1$ in $^3$He.
In this case, the degeneracy temperature below which 
to validate the Fermi liquid theory will be 
severely suppressed
as we approach the transition point $D=0$.
We must be in this situation 
to conclude $D\rightarrow 0$ for finite $n/m$ $(>0)$,
and this was the case of our concern in the previous section.
Nonetheless, in either case, 
the instability does not manifest itself as a thermodynamic phenomenon,
for the vanishing of the Drude weight 
makes sense only in a coherent regime.
Although the case (ii) and its physical consequences
are interesting 
in its own right as a way to destabilize the Fermi liquid state,
here we do not discuss this point any further but
to point out the possible instability. 
%, and
%this point is not discussed any further in the present paper.
In general, % in the presence of Umklapp processes,
(i) and (ii) is to be regarded as independent possibilities,
although these are related with each other  
in an isotropic system, for which $m'=m<\infty$
because of the Landau relation $(1+F^s_1/3)/m^*=1/m$.
In a lattice system, on the other side, 
the relation does not hold so that in principle we can even think of 
the hypothetical system 
where the vertex correction due to $F_1^s$ 
is neglected while the mass $m^*$ is heavily enhanced.
In effect, the effective mass $m'$ is generally anisotropic and
the instability $m'\rightarrow \infty$  %(while $m^*<\infty$) 
may be relevant 
particularly in an anisotropic system, 
{\it e.g.}, in a quasi-two-dimensional system where
Umklapp processes can become quite effective in the direction
perpendicular to a two-dimensional layer.
Then, beyond the instability, if any,
the Fermi surface has to be
rearranged in the way the system avoids the instability.

Finally we shall note that the plasma mode softens as
$\omega_p\propto 1/\sqrt{m'}$ when $m'\rightarrow \infty$:
In a charged system,
the dielectric function $\epsilon(\omega)$
is related to the dynamical conductivity,
\begin{equation}
\epsilon(\omega)=1+\frac{4\pi\mbox i}{\omega}\sigma(\omega).
\end{equation}
Therefore, eq.~(\ref{conductivity}) gives
\[\epsilon(\omega)=1-\frac{4\pi n e^2}{m' \omega^2},\]
from which the plasma frequency is given by
\begin{equation}
\omega_p^2= \frac{4\pi n e^2}{m'}.
\end{equation}
This result is valid only for
\(\omega_p \ll \varepsilon_{k_f}^*\),
since we used the Fermi liquid formula for  $\sigma(\omega)$.

%%%%%%%%%%%%%%%%%%%%%%%%%

In summary, 
we derived a formula to calculate the Drude weight $D$
on the basis of the Fermi liquid theory.
To this end, we considered the state $| p\rangle$ 
which is obtained from the ground state
by boosting it by $p$ in the momentum space.
Thereby we identified 
the instability of the Fermi liquid state 
caused by the metal-insulator `transition' $D\rightarrow 0$.
%It was shown that the ratio $n/m'\propto D$ is calculated directly,
%given the expression for the total energy of the system.
In a lattice system,
electron-electron interactions enhance the effective mass $m'$,
although an unrenormalized mass $m'=m$
is concluded identically in a Galilean invariant system.
As the enhancement is caused by Umklapp processes,
the many-body effect on $m'$ is
largest in the vicinity of half filling $n\sim 1$
in a single-band model.
This was shown for the Hubbard model
in a square lattice by perturbation theory.
As a result, even in a weak-coupling regime,
we obtained the Drude weight $D$
showing the behavior of the `doped insulator'
\({\mbox d}D/{\mbox d}n<0 \) 
in the underdoped region $n\LA 1$. 
It was noted that $m'$ corresponds to the mass
in the quasiparticle current $j_k$
and is generally different from the thermal mass $m^*$
defined through the velocity of quasiparticle $v_k^*$.

\section*{Acknowledgments}
The author would like to thank Professor K.~Yamada
and Dr. K.~Kanki for discussions and 
comments on the Landau Fermi liquid theory,
which were helpful to formulate the results obtained in this article.
It is a pleasure to thank Professor M. Ido for helpful discussions.
This work is supported  by
Research Fellowships of the Japan Society for the
Promotion of Science for Young Scientists.

\appendix
\section{Finite Temperature Formalism}
In this appendix, we outline the derivation of
eq.~(\ref{nm'FL}) by the finite temperature diagram technique.
The formalism used here is based on the work of
\'Eliashberg.~\cite{rf:Eliash}\
According to eq.~(24) of ref.~\citen{rf:Eliash},
the conductivity is given by
\fulltext
\begin{equation}
\sigma_{\mu\nu}(\omega)
%&=&\frac{\mbox i e^2}{2\Omega}\left\{
=\frac{\mbox i e^2}{2\Omega}\left\{
\sum_k%\int\frac{\mbox{d}^3 k}{(2\pi)^3}
v_{k\mu}^*\frac{1}{2T}\frac{\cosh^{-2}(\varepsilon_k^*-\mu/2T)}{
\omega+2\mbox i \gamma^*_k}v_{k'\nu}^*\right.\nonumber\\
%&&+\left.
+\left.
\frac{1}{2}
\sum_{k,k'}%\int\frac{\mbox{d}^3 k \mbox{d}^3 k'}{(2\pi)^6}
z_{k}v_{k\mu}^*\frac{1}{2T}\frac{\cosh^{-2}(\varepsilon_k^*-\mu/2T)
T_{22}(k,k';\omega)}
{(\omega+2\mbox i \gamma^*_k)
(\omega+2\mbox i \gamma^*_{k'})
}z_{k'}v_{k'\nu}^*\right\},
%\end{eqnarray}
\end{equation}
\halftext
where 
\begin{equation}
\gamma^*_k=-z_k\mbox{Im} 
\left.\Sigma(k,\omega)\right|_{\omega=0}.
\end{equation}
For our purpose
we shall use only the real part of $T_{22}(k,k';\omega)$,
for which eq.~(12) of ref.~\citen{rf:Eliash} gives
\fulltext
\begin{equation}
\mbox{Re}T_{22}(k,k';\omega)
=\left(\tanh\frac{\varepsilon^*_{k'}-\mu+\omega}{2T}
-\tanh\frac{\varepsilon^*_{k'}-\mu}{2T}\right)
\mbox{Re}\Gamma(k,k';\omega).
\label{ReT22}
\end{equation}
\halftext
Therefore, noting \[
\frac{1}{4T}\cosh^{-2}\frac{\varepsilon_k^*-\mu}{2T}
\rightarrow
\delta(\mu-\varepsilon^*_{k}), \qquad (T\rightarrow 0)
\]
and $\gamma^*_k\propto T^2 $,
in the collisionless regime $\gamma^*_k/\omega\rightarrow 0$ 
of our concern,
the limit $\omega\rightarrow 0$ leads to
\fulltext
\begin{eqnarray}
\sigma_{\mu\nu}(\omega)
&\rightarrow&\frac{\mbox i e^2}{\omega}
\frac{1}{\Omega}\sum_{k}
\left\{
v^*_{k\mu} v^*_{k\nu}
\delta(\mu-\varepsilon^*_{k})
+\sum_{k'}f(k,k')
v^*_{k\mu}v^*_{k'\nu}\delta(\mu-\varepsilon^*_{k})
\delta(\mu-\varepsilon^*_{k'})
\right\}\\
&=&\frac{\mbox i e^2}{\omega}
\frac{1}{\Omega}\sum_{k}
v^*_{k\mu} j^*_{k\nu}
\delta(\mu-\varepsilon^*_{k}),
\end{eqnarray}
\halftext
where $f(k,k')$ is defined in eq.~(\ref{fkk'}).

In the opposite 
hydrodynamic limit, $\omega/\gamma^*_k\rightarrow 0$,
the imaginary part of $T_{22}(k,k';\omega)$ plays an important role,
and in contrast to the real part, %eq.~(\ref{ReT22}),
to investigate $\mbox{Im}T_{22}(k,k';\omega)$ poses a problem.
This rather complicated task is indispensable, {\it e.g.}, 
to evaluate the dc conductivity $\sigma(0)$ ($ \propto T^{-2}$) of
a clean system at finite temperature.~\cite{rf:YY}\
This point will be discussed in detail 
in the following paper.~\cite{rf:2}


\begin{thebibliography}{99}




\bibitem{rf:Kohn} W. Kohn: Phys. Rev. {\bf 133} (1964) A171.
\bibitem{rf:SS}B. S. Shastry and B. Sutherland:
Phys. Rev. Lett. {\bf 65} (1990) 243.

\bibitem{rf:ELP}V. J. Emery, A. Luther and I. Peschel:
Phys. Rev. {\bf 13} (1976) 1272.
\bibitem{rf:UKO} 
T. Usuki, N. Kawakami and A. Okiji:
Phys. Lett. {\bf 135A} (1989) 476.

\bibitem{rf:1d} H. J. Shultz, in
{\it Correlated Electron Systems,} edited by V. J. Emery
(World Scientific, Singapore, 1992).

\bibitem{rf:KY}K. Kanki and K. Yamada:
J. Phys. Soc. Jpn. {\bf 66} (1997) 1103.

\bibitem{rf:2} T. Okabe: J. Phys. Soc. Jpn. (submitted).

\bibitem{rf:PN}
D. Pines and P. Nozi\`eres:
{\it The Theory of Quantum Liquids I} 
(W. A. Benjamin, Inc., New York, 1966).

\bibitem{rf:comment}
The state $|p\rangle $ will be stable
because a finite lifetime of the state $|p\rangle $
implies a finite dc conductivity at $T=0$.

\bibitem{rf:comment1}
In a homogeneous system,
the relation between $\partial^2 E_p/\partial p^2$ 
and the Drude weight reminds us of 
the derivation of the inequalities $F^s_l/(2l+1)+1>0$
by investigating the stability 
of the Fermi liquid state against an infinitesimal change 
in shape of the Fermi surface,
\(\delta n^s=\sum_{lm}\delta n^s_{lm} 
Y_{lm}(\theta,\phi)\).~\cite{rf:PN}\
Here $\delta n^s$, a spin symmetric part of a deviation 
from the ground-state distribution $n_k^0$, is expanded 
in terms of spherical harmonics $Y_{lm}(\theta,\phi)$ 
for the polar coordinates ($\theta, \phi$) on the Fermi surface.
However, in this argument, 
the case $l=1$ is an exception 
when the parameter $F^s_1$ is canceled out in
the increase of the total energy $\delta E^s_l$ 
caused by the deformation, 
i.e.,  $\delta E^s_l \propto (1+F^s_l/3)/m^*
\propto (1+F^s_l/3)/(1+F^s_1/3)$.
The cancellation  ensures the stability of the Fermi surface
and an unrenormalized Drude weight.
The condition $1+F^s_1/3>0$ is still required for
the effective mass $m^*=(1+F^s_1/3)m$ to be positive definite.

\bibitem{rf:comment3}
It is noted that
the distribution function of noninteracting particles
coincides with the distribution function of quasiparticles
in an interacting system.

  
\bibitem{rf:Noz}P. Nozi\`eres:
{\it Theory of Interacting Fermi Systems}
(Benjamin, New York, 1964).
\bibitem{rf:AGD}A. A. Abrikosov, L. P. Gorkov and I. E.
Dzyaloshinskii:
{\it Methods of Quantum Field Theory in Statistical Physics}
(Pergamon, Oxford, 1965).

\bibitem{rf:com}
Nevertheless, 
the part of $\lim_{\omega\rightarrow 0}\sigma_{\rm inc}(\omega)$
due to quasiparticle scattering is obtained
%expressed in terms of the quantities 
%characterizing quasiparticle
by taking into account the effect of 
the quasiparticle damping.~\cite{rf:2}\


\bibitem{rf:LL}L. D. Landau and E. M. Lifshitz:
{\it Quantum Mechanics}
(Pergamon, Oxford, 1977).

\bibitem{rf:MF} H. Maebashi and H. Fukuyama:
J. Phys. Soc. Jpn. {\bf 66} (1997) 3577.



\bibitem{rf:tJ1} See,
E. Dagotto, A. Moreo, F. Ortolani, D. Poilblanc
and J. Riera: Phys. Rev. {\bf 45} (1992) 10741,
and references therein.
\bibitem{rf:tJ} See, for example, 
A. J. Millis and S. N. Coppersmith:
Phys. Rev. {\bf 42} (1990) 10807;
J. Igarashi and P. Fulde: Phys. Rev. {\bf 48} (1993) 12713.

\bibitem{rf:negD} 
A negative $D$ was observed and discussed in 
the numerical studies of finite systems;
J. Wagner, W. Hanke and D. J. Scalapino:
Phys. Rev. {\bf 43} (1991) 10517;
C. A. Stafford, A. J. Millis and B. S. Shastry:
Phys. Rev. {\bf 43} (1991) 13660.



%\bibitem{rf:Goldstone}J. Goldstone:
%Proc. Roy. Soc. (London), {\bf A239} (1957) 267.

%\bibitem{rf:SST}H. Shiba, R. Shiina and A. Takahashi:
%J. Phys. Soc. Jpn. {\bf 66} (1997) 941.
%\bibitem{rf:Oki}Y. Okimoto, T. Katsufuji, T. Ishikawa,
%A. Urushibara, T. Arima and Y. Tokura:
%Phys. Rev. Lett. {\bf 75} (1995) 109.
%\bibitem{rf:Uru}
%  A. Urushibara, Y. Moritomo, T. Arima, A. Asamitsu, G. Kido and
%  Y. Tokura: Phys. Rev. {\bf B51} (1995) 14103.
\bibitem{rf:Eliash}G. M. \'Eliashberg:
Sov. Phys. JETP {\bf 14} (1962) 886.
%\bibitem{rf:Leggett}A. J. Leggett:
%Phys. Rev. {\bf 140} (1965) A1869.
\bibitem{rf:YY}K. Yamada and K. Yosida:
Prog. Theor. Phys. {\bf 76} (1986) 621.


\end{thebibliography}
\end{document}